\documentclass[conference]{IEEEtran}
\IEEEoverridecommandlockouts
\usepackage{cite}
\usepackage{amsmath,amssymb,amsfonts}
\usepackage{graphicx}
\usepackage{textcomp}
\usepackage{xcolor}
\usepackage{lipsum}
\usepackage{mathtools}
\usepackage{cuted}
\usepackage{amsmath}
\usepackage{bm}
\usepackage{amssymb} 
\usepackage{epsfig} 
\usepackage{subfigure}
\usepackage{textcomp}
\usepackage[utf8]{inputenc}
\usepackage[english]{babel}
\usepackage{lipsum}
\usepackage{mathtools}
\usepackage{cuted}
\usepackage{amsthm}
\usepackage{algorithm}
\usepackage{booktabs}
\usepackage{siunitx}  
\usepackage{algpseudocode}
\usepackage[skip=4pt]{caption}  
\usepackage{mathtools}
\usepackage{siunitx}
\DeclarePairedDelimiterXPP\BigOSI[2]%
  {\mathcal{O}}{(}{)}{}%
  {\SI{#1}{#2}}

\title{Hidden Markov Model Decoding for LDPC Codes}
\author{JC Olivier and Etienne Barnard}
\date{August 2025}
\author{J.C. Olivier and E. Barnard %
\thanks{ J.C. Olivier is with the School of Engineering, University of Tasmania, Hobart, Australia.  Email:  jc.olivier@utas.edu.au}%
\thanks{E. Barnard is with The Faculty of Engineering, North-West University, Potchefstroom, South Africa.  Email: etienne.barnard@gmail.com  }
}
\begin{document}

\maketitle

\begin{abstract}
The paper proposes an iterative Hidden Markov Model (HMM)  for decoding a Low Density Parity Check (LDPC) code.  It is demonstrated that a first-order HMM provides a natural framework for the decoder.  The HMM is time-homogeneous with a fixed transition matrix and is based on a random walk through the encoded frame bits. Each hidden state contains a pair of two encoded bits, and parity checks are naturally incorporated into the observation model.   The paper shows that by implementing a forward-backward smoothing estimator for the hidden states, decoding is efficient and requires only a small number of iterations in most cases.   The results show that the LDPC decoding threshold is significantly improved compared to  belief propagation (BP) on a Tanner graph.  Numerical results are presented showing that LDPC codes under the proposed decoder yield a frame error rate (FER) and decoding threshold comparable to that of a Polar code where Successive Cancellation List (SCL) - Cyclic Redundancy Check (CRC) decoding is deployed. This is shown to be achieved even if the frame length is short (on the order of $512$ bits or less) and a regular LDPC code is used.  
\end{abstract}

\section{Introduction}

Over the last three decades, error correction theory has seen substantial progress, leading to the development of codes that approach channel capacity \cite{new_book,urbancke}.  A notable advance was made in $1993$ when the turbo codes were proposed\footnote{When turbo codes were introduced in the early 1990's, their performance were claimed to approach the Shannon capacity, and that with practical decoding algorithms.  Hence they were initially met with skepticism. However reference C implementation of the decoder was made available to the research community, coinciding with the growing accessibility of the Internet. This allowed independent researchers to replicate the results, leading to rapid validation and acceptance of turbo codes as a breakthrough in coding theory.} \cite{turbo}.  Turbo codes introduced both a novel coding concept and the first practical encoder and decoder capable of operating near the Shannon limit.  

At the same time low density parity check codes (LDPC) were being developed, with a decoder deploying a so-called Tanner graph and belief propagation (BP) \cite{mackay1999good}.  The LDPC codes were in fact proposed during the early 1960's, but the decoder was suboptimal and  the code lay dormant.  During the $1990$'s became clear that turbo codes with its iterative decoding scheme were in fact also based on message passing inference \cite{turbo_BP}.  

Another development for LDPC codes at that time was the so-called spatially coupled LDPC codes, otherwise known as convolutional LDPC codes \cite{conv_ldpc}.  The message passing decoder for these codes approximates a maximum aposteriori probability (MAP) decoder \cite{kudekar}, and the decoding threshold  improved significantly.  These codes are especially effective for long frame length codes, with thresholds that approach Shannon capacity. 

Richardson and Urbancke developed irregular LDPC codes \cite{urbancke}, and these were shown to achieve capacity also.  However at very low Frame Error Rate (FER) levels these codes exhibit an error floor, something that well designed regular codes do not. In general a code with its encoder/decoder pair must yield a FER below $10^{-4}$ before the code can be considered useful.  The energy per bit $E_b/N_0$ required to produce a FER better than $10^{-4}$ is known as the code's \emph{threshold}.  In  general, these codes solved the problem of achieving low threshold $E_b/N_0$ for long frames, but for short frames it remained an unsolved problem. 

A few years after the development of the LDPC codes, Polar codes were developed \cite{polar}.  Initially these codes lacked an optimal decoder and the threshold $E_b/N_0$ was not able to match that of turbo and LDPC codes.  However, an optimal decoder for the Polar codes was proposed by Tal and Vardy  \cite{vardey}, known as the successive‑cancellation list (SCL) - cyclic redundancy check (CRC) aided decoder. The threshold $E_b/N_0$ achieved by the Polar codes  under the SCL-CRC decoder  set a new standard for short frame length codes\footnote{Polar codes have been adopted in the new 5G standard for the control channel.  Control channels often transmit short packets (such as scheduling info and acknowledgments), which require low latency.} --- with excellent  threshold performance in the short frame length  regime, outperforming LDPC and turbo codes.

In this paper new methodology for decoding the LDPC code is proposed, based on a hidden Markov model (HMM).  At the time of writing, the decoder of choice for the LDPC code is the Tanner graph \cite{mackay1999good, turbo_BP}.  There are good reasons why this decoder has become the standard LDPC decoder\footnote{In MATLAB the build-in LDPC decoder function is an implementation of belief propagation (BP) on a Tanner graph.}: 
\begin{enumerate}
    \item Tanner graphs match the \emph{structure} of LDPC codes naturally.
    \item The  sparse parity-check matrix $\mathbf H$  corresponds directly to a bipartite graph.
    \item One set of nodes correspond to the codeword bits (variables), and the other set corresponds to the parity checks (constraints).
    \item An edge exists where $H(i,j) = 1$.
    \item LDPC codes perform message passing (belief propagation (BP)) decoding on the Tanner graph.
    \item If the graph had no cycles, BP would be exact.
    \item For sparse LDPC codes cycles are  long, and  BP is approximate. However empirical results have shown it is remarkably reliable.
\end{enumerate}

In terms of graph theory \cite{friedman}, message passing on a Tanner graph is an inference algorithm on a pairwise Markov random field. This paper demonstrates that the topology of a  hidden Markov model (HMM) also presents a natural framework for the LDPC decoder. The hidden states are bit (variable node) pairs, chosen randomly but must satisfy certain constraints.  The parity-check matrix $\mathbf H$ is  incorporated into the observation model.  The HMM is time-homogeneous and of first order, with a fixed transition matrix. And importantly, hidden state estimation is based on a conventional forward-backward smoothing state estimator\footnote{Low complexity and straightforward to implement in either software or hardware.}.  

The decoded log-likelihood ratio (LLR) of each encoded bit is computed based on the posterior distribution of the hidden states on the HMM chain.  These are then fed back as input for the next iteration. The Markov state chain represents a random walk that reaches all bits (variables) at least once.   

The paper demonstrates that the proposed HMM decoder significantly improves the threshold $E_b/N_0$ of a short LDPC code\footnote{Large frame LDPC codes are not considered in this paper, as they are well understood and existing decoders are near optimal under those conditions.}.  Numerical results are presented that show that the proposed HMM decoder yields a threshold $E_b/N_0$ for regular LDPC codes that is close to the threshold of a Polar code (under SCL-CRC decoding).  This is a new result, and confirms that the proposed HMM decoder for a LDPC code is effective.  

The paper is organized as follows.  Section \ref{hmm} introduces the proposed iterative HMM decoder.  In Section \ref{investigate} the HMM decoder is studied based on an empirical study, and observations are made to motivate the proposed decoding algorithm, described in  Section \ref{algorithm}. Section \ref{results} presents numerical results for two short frame  random regular LDPC codes.  These results include a comparison with a Polar code where a SCL-CRC decoder is used. Conclusions are presented in section \ref{concluded}.  

\section{A time-homogenous hidden Markov model (HMM) decoder for a LDPC code}  \label{hmm}

\begin{figure}
\centering
 \includegraphics[width=0.5\textwidth]{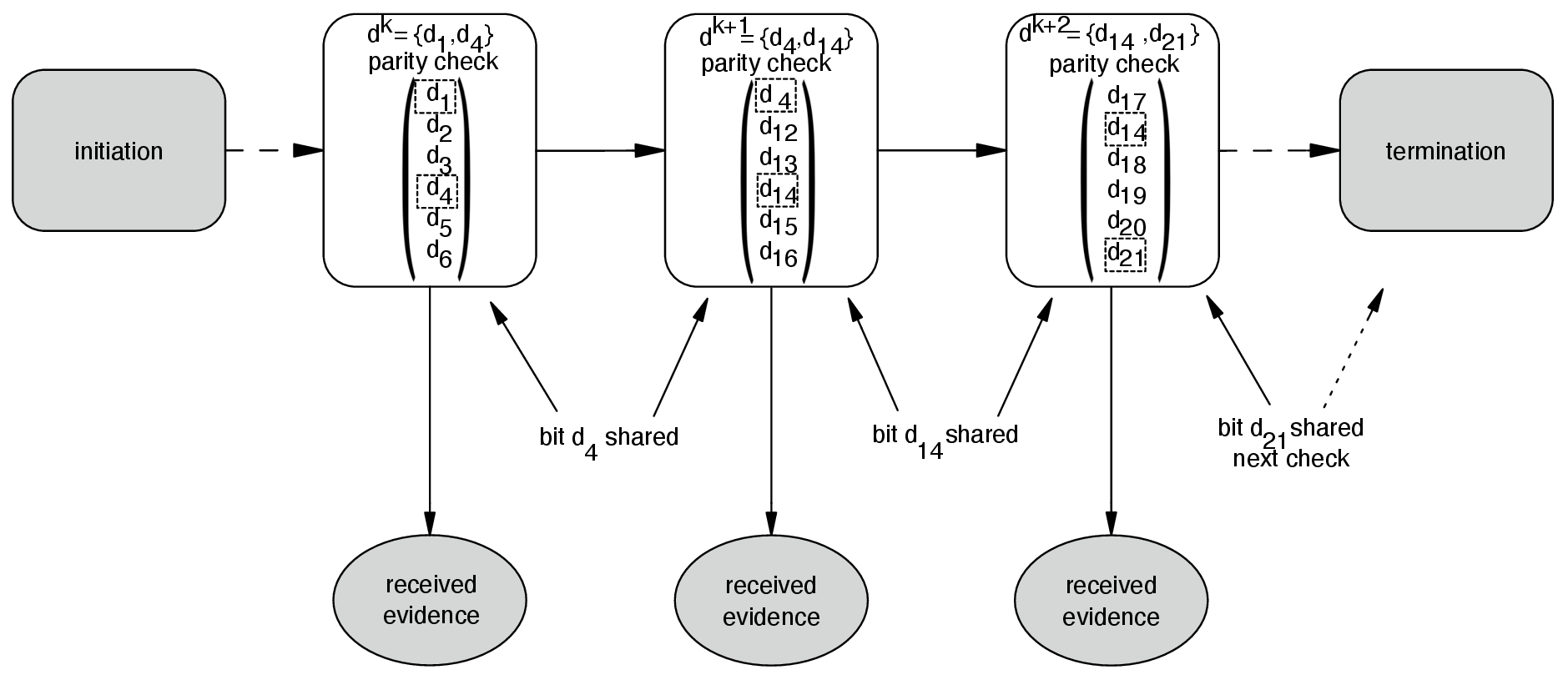}
\caption{The proposed HMM decoder where hidden states are based on two bits chosen randomly from a $6$ bit parity check. However the choice is such that each successive state shares a common bit, which is contained in the same row of the parity check matrix $\mathbf H^\top$. }
\label{hmm_dekodeerder}
\end{figure}

The proposed decoder is based on the fact that the LDPC code consists of $M$ separate constraints. We will represent these constraints by a sequence of \emph{compound} HMM states, where each compound state is associated with one constraint, as shown in Figure \ref{hmm_dekodeerder}. In this first-order HMM each compound state corresponds to particular constraint: two of the encoded bits $(d_i,d_j)$ are treated as observable state indicators, and the four possible values of this pair of bits correspond to the four basic states of the compound state. The constraint consists of $s$ parity check bits ($s=6$ in the figure and below), and the remaining $s-2$ bits in the constraint are treated as latent variables.
In order to link the constraints to one another, we insist that each successive compound state shares one bit with the previous compound state.
This is a constraint on the random selection of the two bits, and ensures that the first-order Markov property of the HMM is satisfied.  The bit that is shared is chosen randomly, but such that the shared bit is present in both  parity checks associated with the states at $k$ and $k+1$.  Hence there are many more compound states than parity checks, and some bits (and constraints) are visited more than once during a random walk through all the bits.  

Note that there are many different HMMs that correspond to a given LDPC code, corresponding to all the ways that a random walk can traverse the constraints and select pairs of state-determinant bits -- under the condition that one shared bit appears in the state-determinant bits of each pair of successive constraints. This multiplicity of options is an important property of our approach, as we show below.

In this paper the encoded frame is assumed to contain $2M$ encoded bits and $M$ uncoded bits, so that the code rate\footnote{The rate can be arbitrary, but in this paper a rate half code is considered.} $R = \frac{1}{2}$. The encoder is a matrix $\mathbf G$, and its structure is given by 
\begin{eqnarray}
    \mathbf G = [\mathbf I ~ \mathbf P].
\end{eqnarray}
Here $\mathbf I$ is the $M$ by $M$ identity matrix, and $\mathbf P$ is the $M$ by $M$ parity  matrix \cite{mackay1999good}.  

Define a parity check matrix $\mathbf H$ which is sparse and regular. The generator matrix \( \mathbf G \) and the parity-check matrix \( \mathbf H \) are orthogonal over \( \mathbb{F}_2 \), meaning
\begin{equation}
\mathbf G \mathbf H^\top = \mathbf 0 \mod 2.
\end{equation} 

For the numerical results to be presented in a later section, each column of $\mathbf H^\top$ has $6$ bits set to $1$, the rest are zero so that the matrix is regular and sparse.  Rows contain $3$ bits that are set to one.   The channel is assumed to be additive white Gaussian, with binary phase shift keying (BPSK) modulation.  The length of the observed (received) noisy symbol frame is $2M$. 

With this arrangement as shown in Figure \ref{hmm_dekodeerder}, and the convention that the four basic states in a compound state are ordered as $(d_m,d_n) = \{(0,0), (0, 1), (1, 0), (1,1)\}$,
the HMM transition matrix between pairs of compound states is given by 
\begin{equation} \label{T}
\mathbf T = \begin{bmatrix}
\frac{1}{2} & \frac{1}{2} & 0 & 0 \\
0 & 0 & \frac{1}{2} &  \frac{1}{2}\\
\frac{1}{2} & \frac{1}{2} & 0 & 0 \\
0 & 0 & \frac{1}{2} &  \frac{1}{2}
\end{bmatrix}
\end{equation}
as the shared bit has to be identical over two successive states. Hence, the HMM theoretically has an infinite settling time. 

\subsection{The emission (observed data) probability $P(\mathbf{y} | \mathbf{d})$ for state $k$}

Consider a six symbol vector $\mathbf y$ observed as
\begin{equation}
\mathbf{y} = (y_1, y_2, y_3, y_4, y_5, y_6).
\end{equation}
These six symbols are the noisy observed BPSK  symbols for the parity check associated with state $k$.  Two of the positions \( i \) and \( j \) (with \( i \ne j \)) correspond to bits \( d_i \) and \( d_j \), representing the current state $(d_i,d_j)$. The remaining four symbols are \emph{latent}, but constrained by the known parity condition for the LDPC code, given by 
\begin{equation}
\mathrm{Check^k} \equiv  d_i \oplus d_j \oplus \bigoplus_{\substack{u = 1 \\ u \ne i,j}}^6 c_u = 0.
\end{equation}
Each check corresponds to a column of matrix $\mathbf H^{\top}$. Here  \( c_u \in \{0,1\} \) are the unknown bit values at the remaining positions.

Each received symbol is independently corrupted by a memoryless channel and additive Gaussian noise, producing a noisy observation \( y_u \). From this, derive the marginal likelihood for each bit value: for each position \( u \), let \( P(y_u | c_u) \) denote the likelihood of observing \( y_u \), given that the underlying bit is \( c_u \in \{0,1\} \).

Then, for a given state \( \mathbf{d}^k = (d_i, d_j) \), the (unnormalized) emission probability is proportional to
\begin{equation}
P(\mathbf{y} | \mathbf{d}) \propto 
\sum_{\substack{c_u \in \{0,1\} \ \\ \text{for}  \ u \ne i,j \\ 
\mathrm{and \, Check^k=0}}} 
\left( \prod_{u \ne i,j} P(y_u | c_u) \right) 
 P(y_i | d_i) P(y_j | d_j). \label{simple}
\end{equation}
This sum is taken over all \( 2^4 = 16 \) possible combinations of the latent bits \( \{c_u\}_{u \ne i,j} \), subject to the parity constraint.

\subsection{Incorporating additional parity checks in state $k$}

In the next section it is shown (based on simulation data) that there are some frames \footnote{A small fraction of the frames if the code is used above its threshold.} that are problematic to decode, and resist decoding based on the low complexity single parity check emission probability presented in the previous section.  

An extended emission probability formulation is presented in this section.  In a later section it will be shown that decoding is performed in stages, depending on how the frame responds to decoding attempts based on the HMM. Frames that resist decoding based on the emissions probability given by (\ref{simple})  are then decoded in a next stage based on the extended emissions probability presented here. 

First of all, note that the matrix $\mathbf H^\top$ contains several parity checks (columns) that contain bits \( d_i \) and \( d_j \), representing the current state $(d_i,d_j)$.   There is a parity check that contains both bits, which was the basis of the emission probability presented in the previous section.  Then there are additional parity checks that involve only one of the bits in  state $(d_i,d_j)$. These are represented as symbol vectors $\mathbf y_{h_v}$ with $v \in \{ 1,2,3,4 \}$, observed (for example) as
\begin{align}
   \mathbf y_{h_1} &= (y_i, y_7, y_8, y_9, y_{10}, y_{11})  \nonumber \\ 
   \mathbf y_{h_2} &= (y_i, y_{12}, y_{13}, y_{14}, y_{15}, y_{16})  \nonumber \\
   \mathbf y_{h_3} &= (y_j, y_{17}, y_{18}, y_{19}, y_{20}, y_{21})  \nonumber \\
   \mathbf y_{h_4} &= (y_j, y_{22}, y_{23}, y_{24}, y_{25}, y_{26}).
\end{align}
Here $(y_i,y_j)$ are the noisy observations corresponding to $(d_i,d_j)$, and  $y_7,\cdots,y_{26}$ represent latent noisy variables. The positions \( i \) and \( j \) (with \( i \ne j \)) correspond to bits \( d_i \) and \( d_j \) of the state $(d_i,d_j)$. The remaining latent symbols are constrained by the known parity-check conditions for the LDPC code, given by 
\begin{align}
\mathrm{Check^k_{h_\alpha}} &\equiv  d_v \oplus  \bigoplus_{\substack{u = 1 \\ u \ne v}}^6 c_u = 0, \alpha \in \{1,\cdots,4 \} \, v\in\{i,j \}.
\end{align}
Each check corresponds to a column of matrix $\mathbf H^{\top}$. Here  \( c_u \in \{0,1\} \) are the unknown bit values at the remaining positions, corresponding to the latent variables in $\mathbf y_{h_\alpha}$.

Each received symbol is independently corrupted by a memoryless channel and additive Gaussian noise, producing a noisy observation \( y_u \). From this, derive the marginal likelihood for each bit value: for each position \( u \), let \( P(y_u | c_u) \) denote the likelihood of observing \( y_u \), given that the underlying bit is \( c_u \in \{0,1\} \).

Then, for a given state \( \mathbf{d}^k = (d_i, d_j) \), the (unnormalized) emission probability is proportional to
\begin{align}
P(\mathbf{y} | \mathbf{d}) \propto &
\sum_{\substack{c_u \in \{0,1\} \ \\ \text{for}  \ u \ne i,j \\ 
\mathrm{ Check^k=0}}} 
\left( \prod_{u \ne i,j} P(y_u | c_u) \right) 
 P(y_i | d_i) P(y_j | d_j) ~\times \nonumber \\
& { \prod_{\ell\in\{1,2\}}} \left \{ \sum_{\substack{c_u \in \{0,1\} \ \\ \text{for}  \ u \ne i \\ 
\mathrm{ Check^k_{h_\ell}=0}}} 
\left( \prod_{u \ne i} P(y_u | c_u) \right) 
 P(y_i | d_i) \right \} ~ \times \nonumber \\
&  {\prod_{\ell\in\{3,4\}}} \left \{ \sum_{\substack{c_u \in \{0,1\} \ \\ \text{for}  \ u \ne j \\ 
\mathrm{ Check^k_{h_\ell}=0}}} 
\left( \prod_{u \ne j} P(y_u | c_u) \right) 
 P(y_j | d_j) \right \}. 
  \label{uber}
\end{align}
The sums are taken over all possible combinations of the latent bits, subject to the parity constraints.  This expression for the emission probability of hidden state $k$ uses  all available information (for state $k$), within the general framework of the first-order HMM shown in Figure \ref{hmm_dekodeerder}.

\subsection{The forward-backward smoothing estimator, and subsequent iterations on the HMM chain}

Since there are four possible values of the state $\mathbf{d}^k$, the emission probability is computed for each of the four combinations of \( (d_i, d_j) \). The resulting conditional probabilities are then used in the HMM forward-backward smoothing algorithm \cite{hmm_paper1,hmm_paper2}, which yields the posterior distributions for the hidden states --- the algorithm is provided in the Appendix. Note that the inference algorithm on the HMM chain is a message passing scheme, and has a straightforward and low-complexity implementation (in software or hardware).    

With all $2M$ LLR estimates available after completing the forward-backward inference (one iteration), these estimates now update the LLR vector $\mathbf y$ for the next iteration, assuming the frame did not yet decode. A small number of iterations ($5$ were used during simulations) are performed based on a fixed random walk.  This is explained in more detail in the next section.  If decoding was  unsuccessful, a new random walk is generated, and the process is repeated.   This procedure is terminated if decoding is successful, or if the maximum number of iterations allowed is exceeded. 

As the observed data for successive iterations are  based on the estimated LLR from previous iterations, it is clear that decision feedback is performed -- with the ever-present effects of error propagation.   The hope is that each iteration will yield an improved  posterior estimate, and that an equilibrium will be achieved after a finite number of iterations -- one that decodes the received noisy frame.    

In the next section it is shown that this is indeed the case for many frames, but there are frames that do not decode under these conditions. The paper proposes additional strategies for mitigating frames that are difficult to decode, so that the decoding algorithm proposed in Section \ref{algorithm} contains multiple stages. 

During simulation studies it was observed that frames that did not decode under a Tanner graph, nor under an HMM utilizing the emission probability given by (\ref{simple}), often did decode under an HMM decoder when utilizing the emission probability given by (\ref{uber}).  Thus, it has been experimentally verified that the HMM presents a more effective iterative decoder for a LDPC code compared to the Tanner graph under BP.  

\section{Decoding behavior of the iterative HMM LDPC decoder} \label{investigate}



The HMM chain under a random walk has states that repeat, a consequence of the need to visit each bit (variable) at least once.  All messages reach all states (due to the infinite settling time), but the repeated states have an edge over unrepeated states --- their state distribution is estimated and propagated along the chain more than once.  

To demonstrate that the repeated hidden states have a significant effect on the HMM behavior under iterative decoding, an experiment was performed where the repeated states were disabled.  This was accomplished by having the evidence of the repeated states set to uncertainty (LLR set to zero).  With this modification, those states pass on any message without modifying it. 

By removing the additive white Gaussian noise, and placing wrong symbols at a number of random positions in the frame,  frames were selected that did not decode under Tanner graph decoding. On those frames the iterative HMM decoder was deployed, with and without the repeated states.  The results revealed that when the repeated states were active, the iterative HMM decoder 
\begin{enumerate}
    \item Converged to an equilibrium significantly faster.
    \item  Often moved towards an equilibrium with significantly fewer errors. 
\end{enumerate}

\begin{figure}
\centering
 \includegraphics[width=0.53\textwidth]{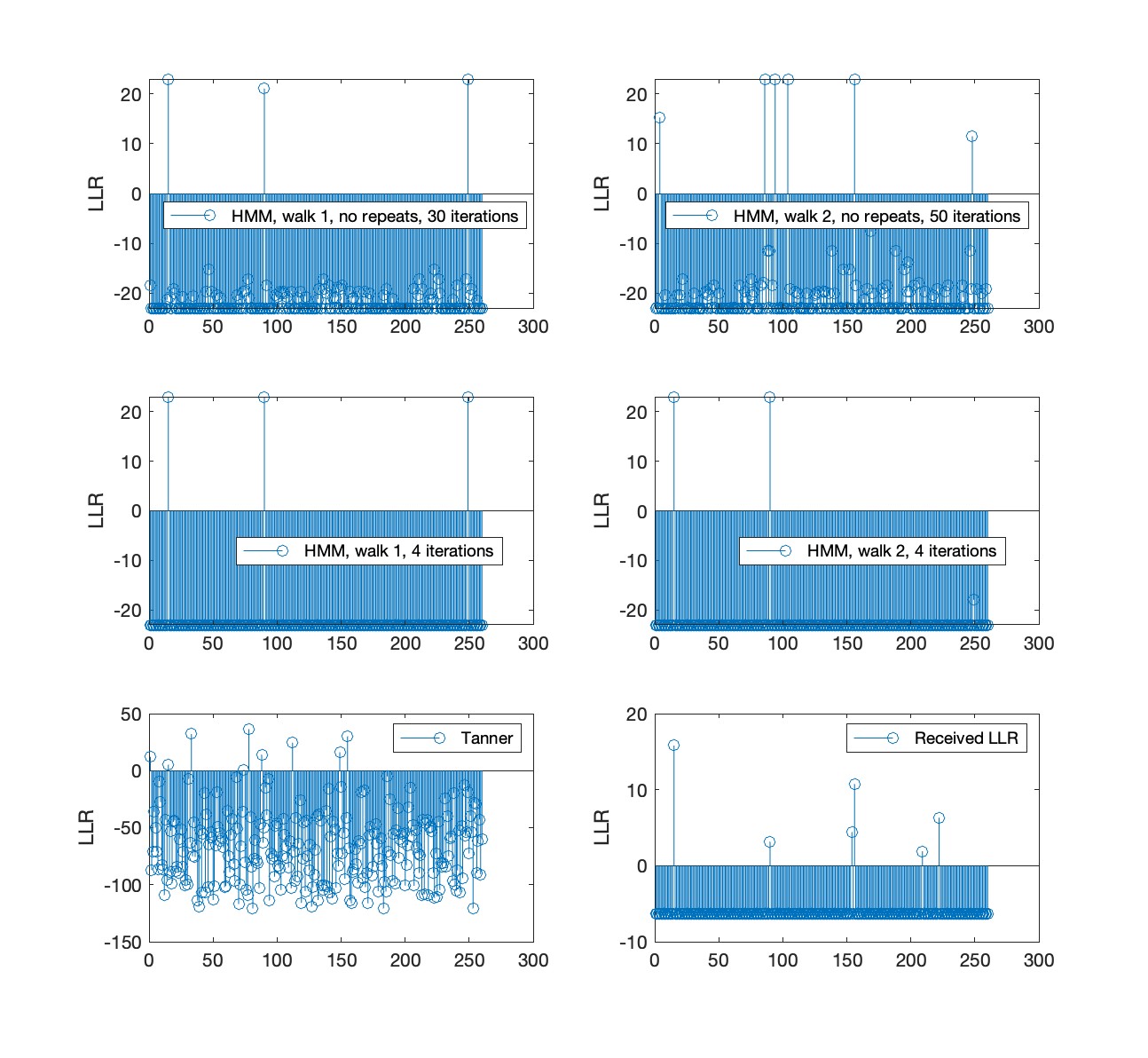}
\caption{A frame where the iterative HMM decoder were deployed with and without the repeated states (disabled by modifying the evidence for the repeat states).  }
\label{repeat_states}
\end{figure}

Figure \ref{repeat_states} shows two cases: 1) the equilibrium state was the same with and without repeat states, and 2) the equilibrium state errors were significantly less when the repeat states were active.  Regardless of the number of errors in the equilibrium state, the iterative HMM decoder converges in a small number of steps when the repeat states are active. This evidence seems to support the ideas presented above. 

Thus, a likely explanation for the observed outcome in Figure \ref{repeat_states}, is that for some random walks the repeated states avoid catastrophic LLR values.  The better LLR values sampled by these repeated hidden states, along with their estimated distribution, are then repeated along the chain through message passing, leading to an improved outcome.  This mechanism is random, and implies a need for multiple  iterative HMM decoding attempts, each based on a different random walk. This strategy was used in the numerical results presented in Section \ref{results}.  

In the next subsection an example of a frame that decoded rapidly, and one that resisted decoding even after a large number of random walks were selected (a maximum of $100$ random walks were used in this paper).  


\subsection{Class 1: frames that decode rapidly  under the HMM decoder}

In the waterfall region of the code and specifically in the threshold region where the code is considered useful, the frame error rate (FER) is required to be small, typically less than $10^{-4}$.  In the threshold region, a large percentage of frames decode rapidly under the HMM decoder\footnote{Section \ref{results} presents statistics to demonstrate this fact.}. 

\begin{figure}
\centering
 \includegraphics[width=0.52\textwidth]{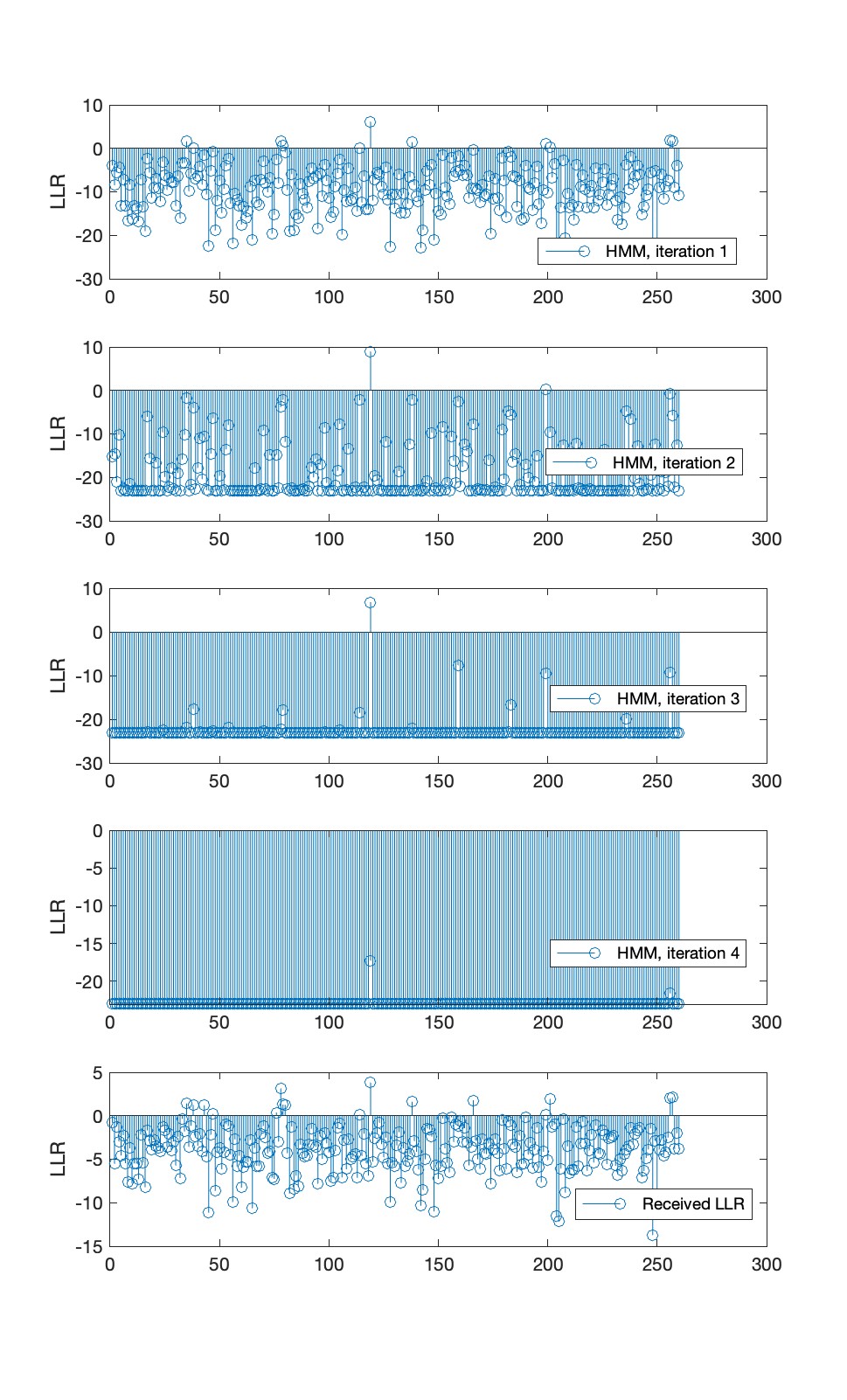}
\caption{The iterative HMM applied to an AWGN received frame that decoded successfully.  The frame also decoded  under a Tanner graph with flooding. The transmitted frame contained the all zero codeword.  The frame was $260$ bits long and $E_b/N_0 = 2.75$ dB.  }
\label{did_decode}
\end{figure}

To make this point clear, consider a frame with $260$ bits, where the all-zero codeword was transmitted. The channel corrupts the symbols based on white additive Gaussian noise (AWGN).  An example received frame is shown in Figure \ref{did_decode}.  The Tanner graph decoded the frame correctly,  as did the HMM decoder after four iterations -- but the first two random walks did not decode (each after $5$ iterations).   The received LLR values are shown in the bottom subplot, and the all zero codeword was transmitted.  These results are based on the low complexity emission probability formulation given by (\ref{simple}). 

\begin{figure}
\centering
 \includegraphics[width=0.52\textwidth]{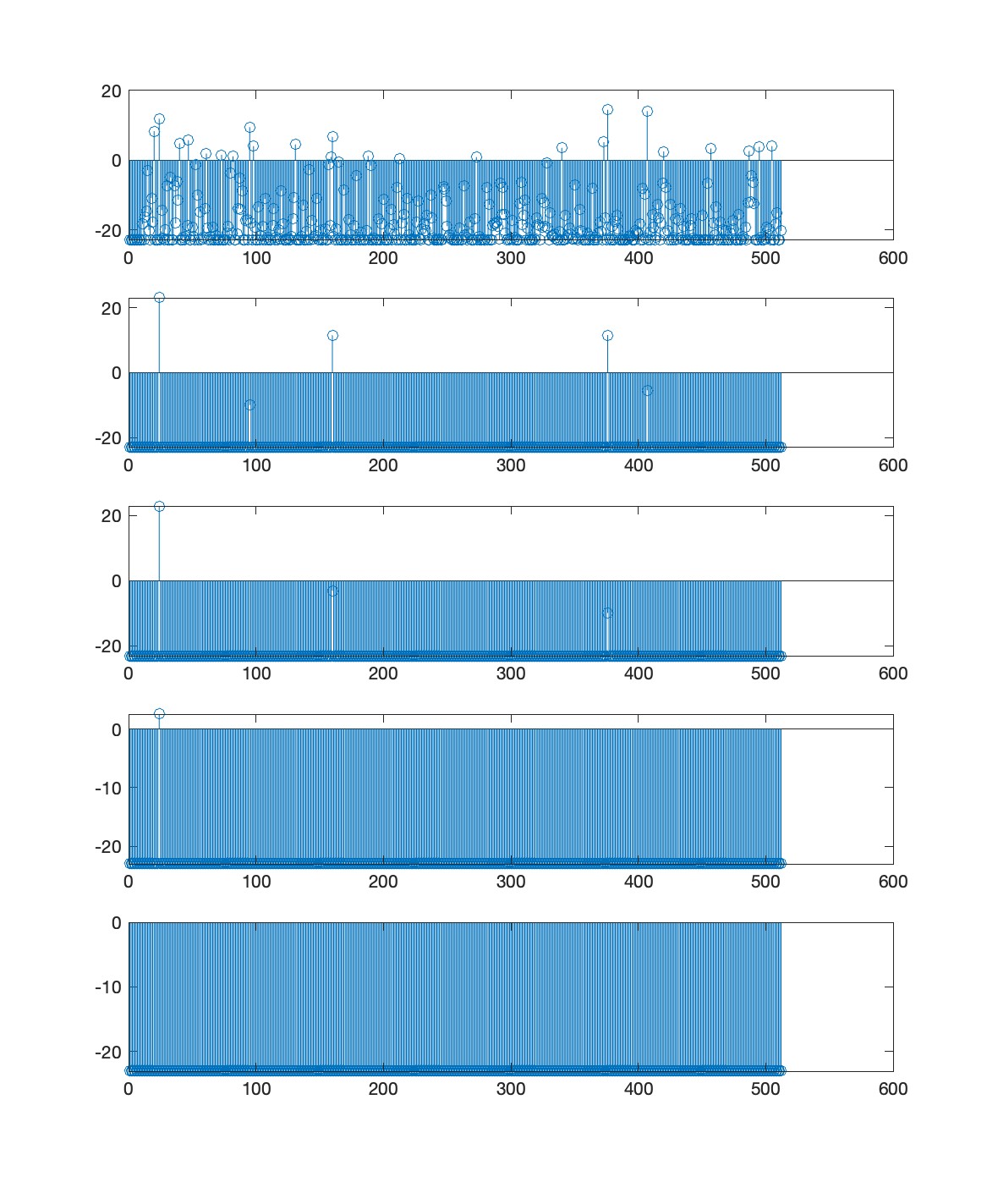}
\caption{Five iterations of the HMM decoder where the extended emission probability (\ref{uber}) was utilized.  A frame with $512$ bits was received in an AWGN channel.   The frame did not decode under the HMM decoder with the low complexity emission probability given by (\ref{simple}), nor did it decode under a Tanner graph and BP. The transmitted frame contained the all zero codeword.   }
\label{demo1}
\end{figure}

Some frames decode rapidly under the HMM decoder and the extended emission probability given by (\ref{uber}), but the Tanner graph under BP is unable to decode the frame. This demonstrates that the HMM decoder is a more effective decoder than the Tanner graph under BP.  Figure \ref{demo1} shows such a frame.  It decoded rapidly under the HMM decoder and (\ref{uber}), but defeated both the HMM under (\ref{simple}) and the Tanner graph under BP. 

\subsection{Class 2: frames that are difficult to decode, even  after a large number of random walks were utilized}

There are frames that do not decode, even after $100$ random walks were utilized. Not surprisingly, these frames also do not decode under a Tanner graph with BP.  This paper proposes two strategies that will decode these frames in many cases. These strategies improve the LDPC code's threshold significantly, and  are presented below. They also form the basis for the staged HMM decoding algorithm will be presented in a section to follow.  

\subsubsection{Strategy 1: combine the two decoders}

\begin{figure}
\centering
 \includegraphics[width=0.5\textwidth]{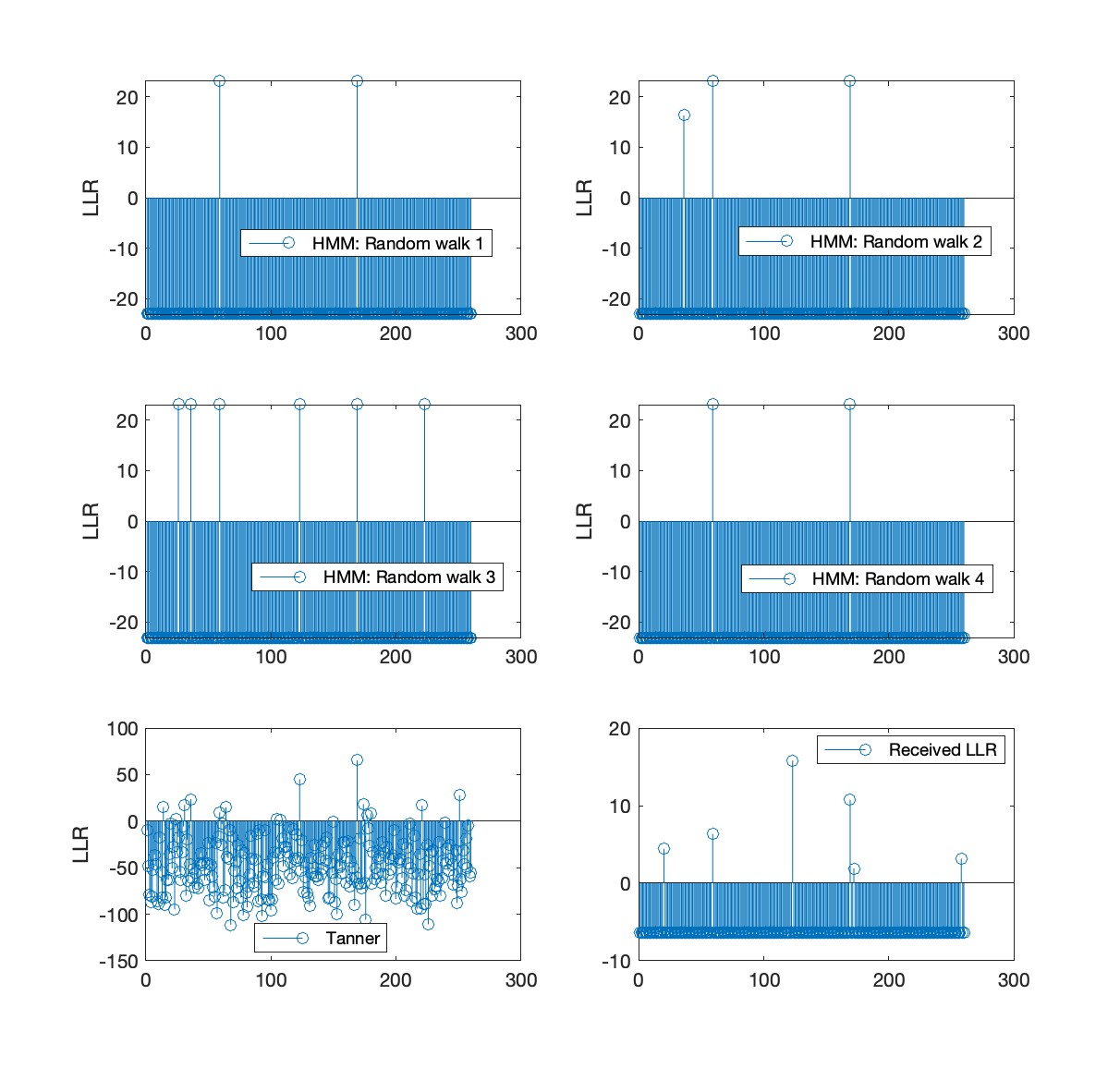}
\caption{A frame that did not decode even after  $100$ random walks.  The received LLR values are shown at the bottom right, and the Tanner graph equilibrium in the lower left subplot.  Four HMM decoding attempts are shown, each with a fixed (but different) random walk, with $5$ iterations for each random walk. }
\label{not_decode}
\end{figure}

To demonstrate this strategy, consider the artificial received frame shown in Figure \ref{not_decode}. 
Here the additive white Gauss noise was removed, and several randomly positioned received LLR values were chosen to be positive (wrong).  Note that the Tanner graph ended up in a state where there are many bits decoded in error. And not one of the $100$ random walk iterative HMM decoding attempts (based on (\ref{simple})) was successful either. 

However, note that some of the random walks yielded equilibrium states where there are only a small number of errors that remain. Can these be corrected somehow?  Yes, by passing the HMM \emph{output} LLR as an \emph{input} to an additional Tanner graph attempt (under BP).  As the HMM was able to correct so many of the errors, the additional Tanner graph stage cleans up the remaining few errors.  

The Tanner graph is unable to decode the frame on its own, but after a random walk of the HMM decoder was able to remove many wrong (and possibly a few catastrophic) LLR values, the Tanner graph is able to correct the remaining problematic LLR values, and decode the frame.  To demonstrate this idea, the example in Figure \ref{not_decode} contains one catastrophic LLR value, in this case the LLR was set to $+16$ dB.  Some of the HMM random walks were able to remove it and not introduce too many other bit errors during the process -- that was the key step in being able to decode this frame.  

Although this frame was artificially created to demonstrate this effect, on the basis of simulations, it was verified that there are frames under an AWGN channel that also exhibit this effect. These frames are difficult to decode and resist both the HMM decoder and the Tanner decoder when used in isolation.  But after one of the random walks corrects critical  incorrect LLR values, the Tanner graph is able to clean up remaining errors.  

A number of problematic frames can be decoded by combining the two decoders --- making the most of having two different decoders. This strategy improves the LDPC code's decoding threshold significantly. 

\subsubsection{Strategy 2: frames that do not decode after strategy $1$ was deployed -- statistical analysis required}

There are other problematic frames, those that do not decode after strategy $1$ was utilized.  During simulation studies performed to study the HMM decoder's iterative response to these problematic frames, it was observed that \emph{first iteration} LLR obtained after a single forward-backward HMM estimator (based on different random walks), contain bit reliability information. 

\begin{figure}
\centering
 \includegraphics[width=0.52\textwidth]{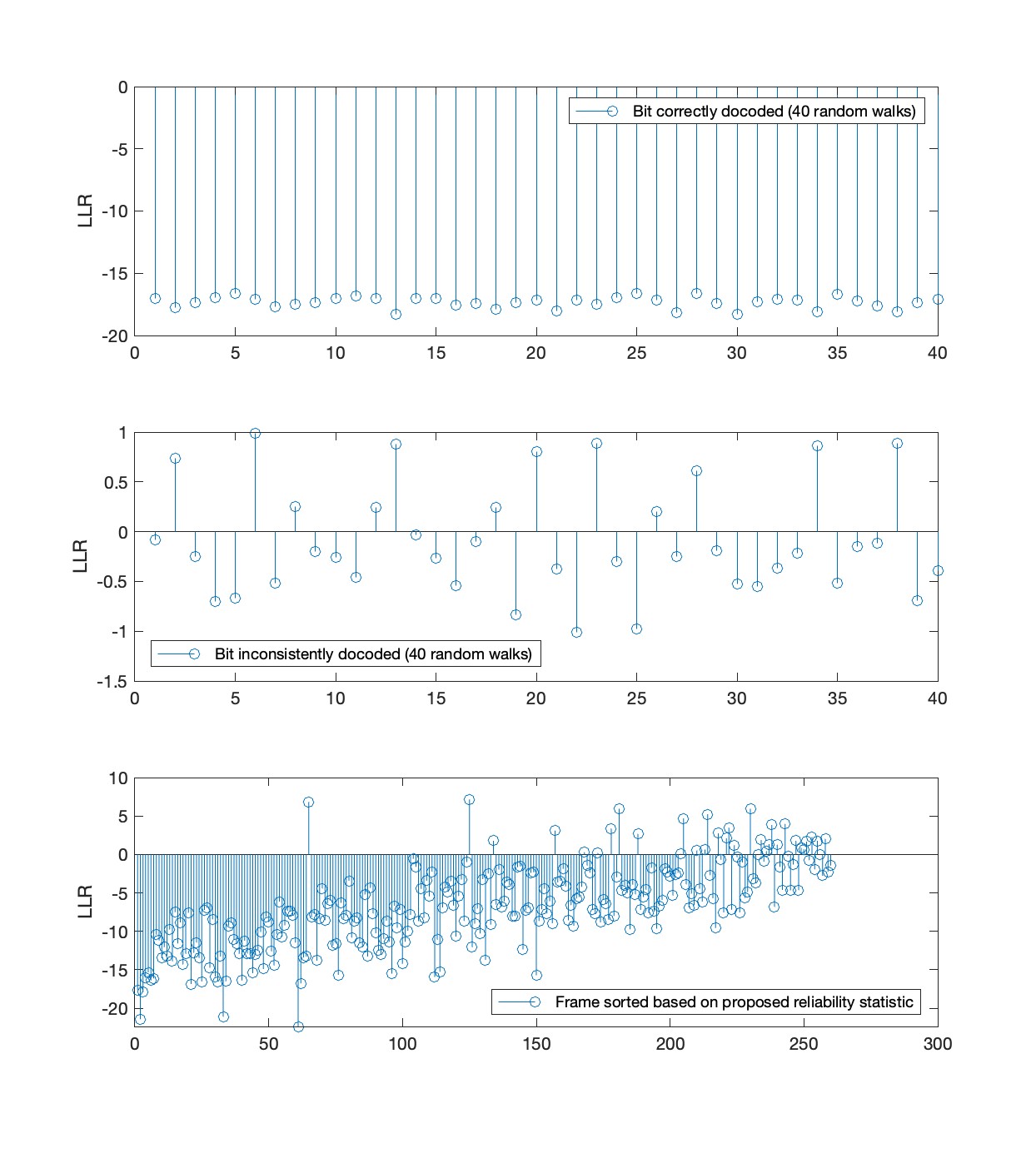}
\caption{The behavior of a bit reliably decoded versus a bit that is unreliable for $40$ HMM first iterations. The last subplot shows the sorted frame of $260$ bits.  Note that many of final bits in the sorted frame were indeed wrong, which is obvious in this case as the transmitted frame was the all zero codeword.}
\label{reliability}
\end{figure}

Consider an example of this class of frame shown in Figure \ref{reliability}. The estimated LLR is shown for two bits, and were obtained after $40$ random walks were completed -- each with one forward-backward iteration on the chain.  The top plot shows a bit correctly decoded over the $40$ random walks, and the other a bit incorrectly decoded.  The incorrectly decoded bit is inconsistent, while the correctly decoded bit is consistent -- and this situation is typical.  This information can be exploited to our advantage, as shown below. 

Let $\mathbf x$ denote a vector containing the first iteration LLR estimates for bit $i$, after a number of random walks were completed (each with one iteration). It is proposed to compute a reliability statistic for bit $i$ denoted  $\gamma_i$, given by 
\begin{eqnarray}
  \gamma_i =   \frac{\mathrm{std}(\mathbf x)}{\mathrm{mean}(\mathbf x)}. \label{metric}
\end{eqnarray}
The frame with LLR values sorted based on the proposed statistic $\gamma$, is shown in Figure \ref{reliability} at the bottom.  In this case where the all zero codeword was transmitted ($260$ bits), it is clear that the last part of the sorted frame contains many incorrectly decoded bits.  Note that the sorting is not based on the magnitude of the LLR values, it is based on the  reliability of each LLR value, as assessed by (\ref{metric}). 

\begin{figure}
\centering
 \includegraphics[width=0.52\textwidth]{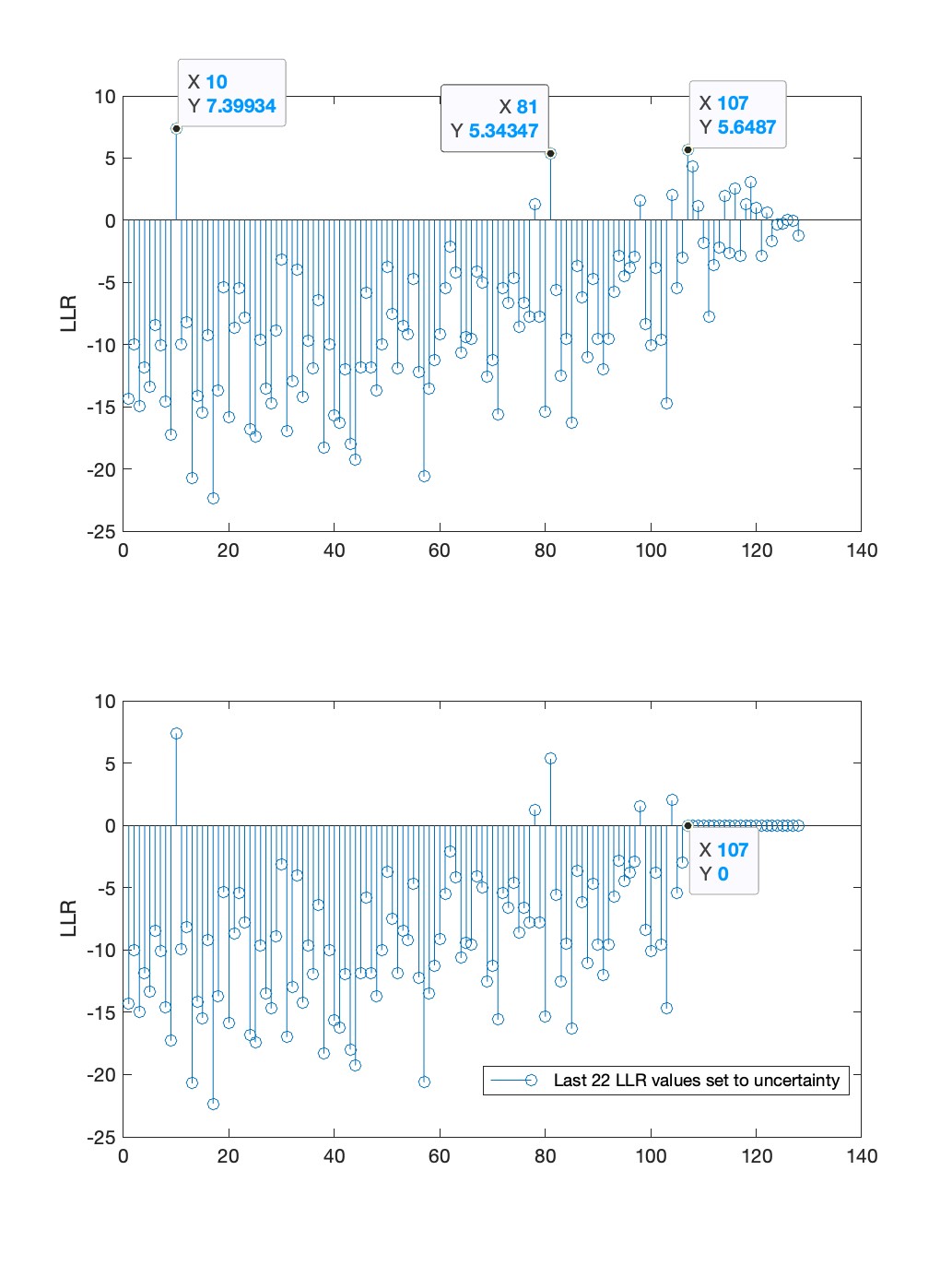}
\caption{A problematic frame that resisted decoding with a HMM and a Tanner graph on an AWGN channel.  After it was sorted and the last $22$ symbols set to zero, it decoded when unsorted.} 
\label{stats_example}
\end{figure}

The LLR values of a small number of bits starting at the last bit can be set to zero.  Then the sorting can be reversed, and the HMM iterative decoder can be applied again.
To illustrate this approach with an example, Figure \ref{stats_example} shows a sorted $128$ bit frame. This frame resisted all decoding efforts, however when the last $22$ LLR's of the sorted frame were declared unreliable as shown, and then unsorted, it decoded.  In spite of the remaining large incorrect LLR values at $10$ and $81$, the decoder easily decoded this frame.  This strategy improves the decoding threshold significantly. This is accomplished in spite of also removing the LLR of some bits that were correctly decoded -- overall there is a net gain. 

During MATLAB simulations, LLR values were sorted, and then the unreliable symbols were  set to zero in increments of $2 \%$, up to $20 \%$ of the frame length. 

 \section{Proposed HMM decoder: multistage decoding algorithm} \label{algorithm}

The proposed HMM decoding algorithm is given in Algorithm $1$ below.
\begin{algorithm} 
\caption{Proposed HMM decoder for a LDPC code}
\begin{algorithmic}
 \State \textbf{Input:} Received frame with LLR values. 
\State \textbf{Output:} Decoded information bit sequence $\mathbf d$.\\
\State \textbf{Stage 1:} Attempt decoding based on the iterative HMM decoder with emissions probability given by (\ref{simple}). Utilize a  number of different random walks, each with a small number of iterations.  If HMM iterative decoding fails, pass the HMM output LLR values to the Tanner graph as an input, and attempt decoding the frame.  If decoding is a success, terminate the decoding algorithm. 
\State \textbf{Stage 2:} If stage 1 failed, attempt decoding based on the iterative HMM decoder, but based on the extended emissions probability given by (\ref{uber}). Utilize a number of different random walks, each with a small number of iterations. If HMM iterative decoding fails, pass the HMM output LLR values to the Tanner graph as an input.  If decoding is a success, terminate the decoding algorithm. 
\State \textbf{Stage 3:} If Stage $2$ failed, perform statistical sorting, then set a small number of the unreliable bits to zero.  Unsort and attempt decoding based on the HMM iterative decoder with emissions probability given by (\ref{simple}).  If decoding is unsuccessful, feed the HMM output to a Tanner graph with flooding/BP and attempt decoding.  Terminate the decoding algorithm if decoding was a success. 
\State \textbf{Stage 4:}  If stage $3$ failed to decode, sort the frame LLR values based on the reliability metrics, but now utilizing the extended emissions probability given by (\ref{uber}).  Set a small number of the unreliable LLR values to zero. Now un-sort and  apply the iterative HMM decoder.  If required, also feed the output of the HMM to a Tanner graph with flooding/BP and attempt decoding.  Terminate the decoding algorithm if decoding was a success. 
\State \textbf{Stage 5:} If stages $1$ to $4$ failed, declare the frame to be incorrectly decoded. Terminate the decoding process.  
\end{algorithmic}
\end{algorithm}

The computational complexity of the proposed algorithm is low, as the successive stages can all be performed by reusing information already computed during previous stages.  For example, in simulations performed in the next section, $100$ random walks were utilized, each with $5$ iterations on the HMM chain.  Thus for a frame that did not decode during stage $1$, all the reliability information was in fact generated already, and kept in a matrix for later use in case it was needed.

Secondly, problematic frames that require all the stages of the proposed HMM decoder are rare, only a small fraction of the decoded frames end up in stage $4$.  


\section{Numerical results} \label{results}

Two short frames are considered in this section --- $128$ and $512$ bits long.  For these frame lengths, results from the literature where a Polar code was utilized were obtained. Thus a comparison of the FER results for a regular LDPC code under the proposed HMM decoder, and that of state-of-the-art Polar codes were performed.  The FER obtained by utilizing the Tanner graph with BP is also shown for both cases.

During all simulations performed in this paper, a frame was deemed decoded if there were $2$ or fewer bits in error.  This is because in practice it is straightforward to correct $2$ or fewer errors with an LDPC code\footnote{To check for $2$ bits in error, enumerate all $2$ bit corrections, a process that requires $$\binom{M}{2} = \frac{M(M-1)}{2} \approx \mathcal{O}(M^2)$$ encoding operations. The correct uncoded bit string of $M$ bits will yield encoded bits that are close to the decoded parity bits. This is only an option for short frames where $M^2$ is a reasonable number. Also note that encoding operations now only involve bits that are modified.}.

The settings for the simulations presented were as follows:
\begin{enumerate}
\item For each frame a different random binary data vector was generated.
\item The Tanner graph used BP and flooding with $250$ iterations. 
\item The HMM decoder used a maximum of $100$ random walks. 
\item Each random walk utilized $5$ forward-backward iterations on the HMM chain. 
\item For stages $3$ and $4$, unreliable LLR values were set to zero in increments of $2 \%$, up to $20 \%$ of the frame length.
\item For the $512$ bit Polar code, the effective rate is $0.484375$ and for the $128$ bit Polar code it is $0.46875$. For the LDPC code the rate is $0.5$, and thus a small adjustment was made to the FER plots. 
\end{enumerate}
These are conservative settings, yielding a low complexity HMM decoder\footnote{Compared to a Polar code where a List-CRC decoder is utilized.}.  Consequently, the FER for the HMM decoder presented can be improved by increasing the number of random walks and the iterations per walk -- at the cost of introducing more complexity.  

\subsection{FER and threshold results for a $512$ bit Polar and regular LDPC code}

\begin{figure}
\centering
 \includegraphics[width=0.52\textwidth]{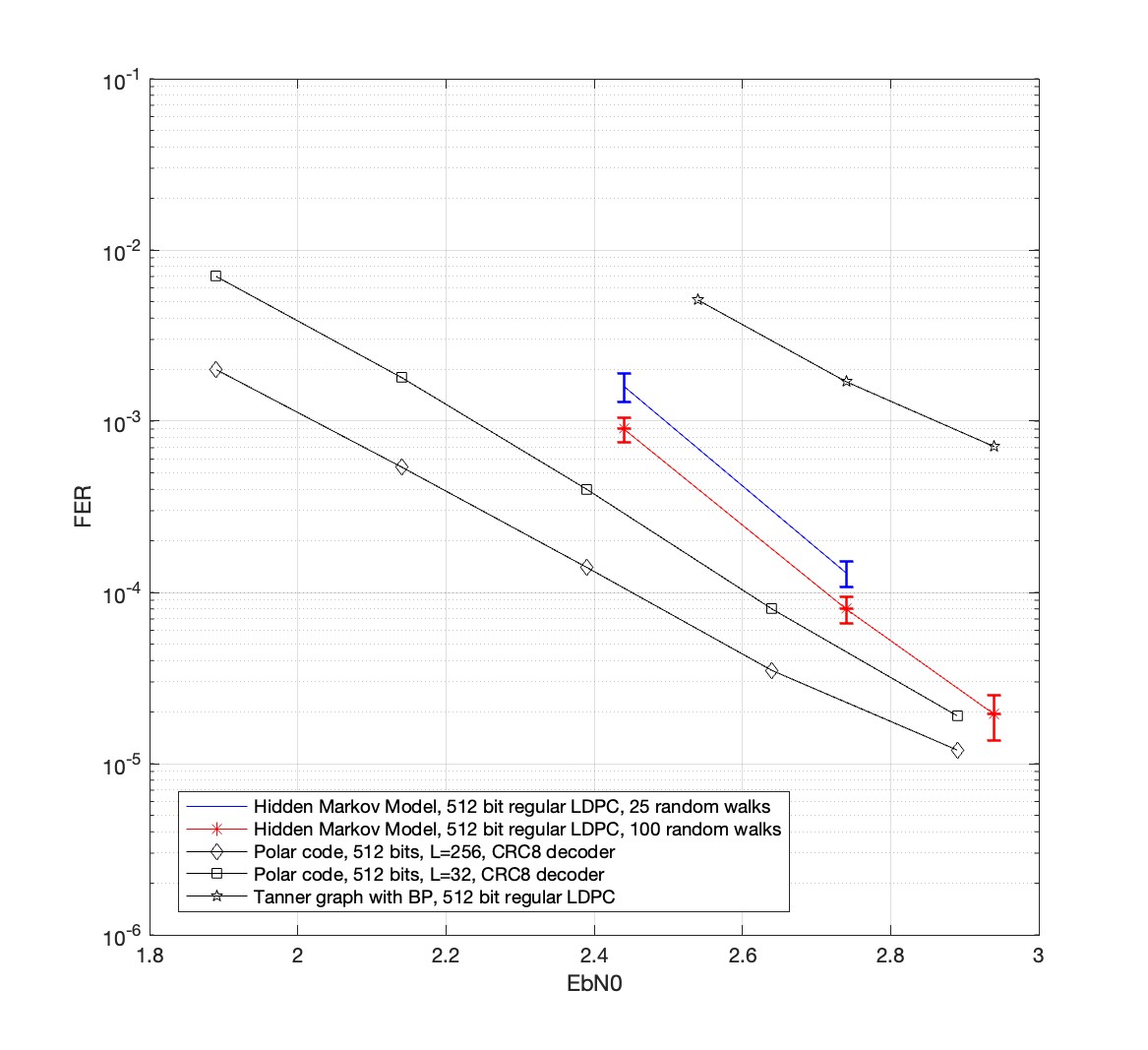}
\caption{The FER performance for a LDPC code under the proposed HMM decoder,  compared to a Polar code under a list-CRC decoder for a $512$ bit frame.  The Polar code results were obtained from Tal and Vardy \cite{vardey}.   }
\label{polar}
\end{figure}

In \cite{vardey} results for a $512$ bit Polar code with rate half was presented.  The decoder was a list decoder aided by a CRC check, and  optimal for the Polar code.   The FER for a $512$ bit regular LDPC code under the proposed decoder, as well as the FER for the Polar code with $L=32$ and $L=256$, is shown in Figure \ref{polar}. The HMM FER is shown for $25$ and $100$ random walks.   

It should be noted that the proposed HMM decoder is based on a forward-backward message passing estimator on the Markov chain, which is  low complexity and straightforward to implement in software or hardware.  The Polar code under a List-CRC decoder yields a gain in threshold, but is very complex and requires significant resources.  Thus when selecting a code, the system designer has to consider the complexity of implementing a List-CRC decoder given the modest improvement in threshold over the low-complexity LDPC code. Furthermore, the HMM decoder was implemented based on $100$ random walks, and $5$ iterations per walk.  These settings can be increased, and the LDPC threshold will then improve, at the cost of additional complexity. 

The performance of LDPC codes under belief propagation decoding on a Tanner graph is limited by the presence of loops and trapping sets \cite{urbancke}. Consequently, the FER response when a Tanner graph is deployed as a decoder tends to underestimate the true potential of the LDPC code.  The HMM decoder yields a significantly improved threshold over the Tanner graph under BP, and the gradient of the FER plot is also improved.  The gradient of the FER versus energy-per-bit plot is a proxy for how well a decoder is exploiting the redundancy introduced by the encoder.  Thus it is clear that based on the proposed HMM decoder, the full performance of the underlying regular LDPC code is better revealed. 

 Table \ref{tabel2} shows how the different stages were utilized, for a simulation at $2.7$ dB.  In this case, stage $2$ was turned off, so all failed frames from stage $1$ were passed to stage $3$. This demonstrates that the majority of frames that can be decoded through statistical analysis is decoded in stage 3, where the low complexity emission  probability is utilized.  Stage four requires the extended emission probability, and decoded $7$ of the $9$ frames that defeated stage $3$.
\begin{table}[h!]
\centering
\caption{HMM Stage Requirement for a $512$ bit code at $E_b/N_0 = 2.7$ dB.  }
\label{tab:decoding-results}
\begin{tabular}{@{}lS[table-format=6.0]@{}}
\toprule
\textbf{Decoding Outcome} & \textbf{Frame Count} \\
\midrule
Decoded in Stage 1        & 139641 \\
Decoded in Stage 3        & 350 \\
Decoded in Stage 4        & 7 \\
Failed to Decode          & 2 \\
\midrule
\textbf{Total Simulated Frames} & 140000 \\
\bottomrule
\end{tabular}
\label{tabel2}
\end{table}



\subsection{FER results for a $128$ bit Polar and regular LDPC code}

In \cite{cioffi} sophisticated  methodology for designing Polar codes was proposed.  The so-called Iterative Message-Passing (IMP) algorithm uses machine learning in the design of the initialization operations, the update operations, and the post-processing phase.  The Polar code is also tuned to the channel, which includes the signal-to-noise ratio.

The FER results presented in \cite{cioffi} include several code designs, but in this paper the $128$ bit frame Polar code presented in Figure $5$ of \cite{cioffi} was selected. The results showed FER for list lengths $L \in \{2,4,8\}$, and the decoder is aided by a CRC code.  In this paper only the $L=8$ FER results are shown. 

\begin{figure}
\centering
 \includegraphics[width=0.525\textwidth]{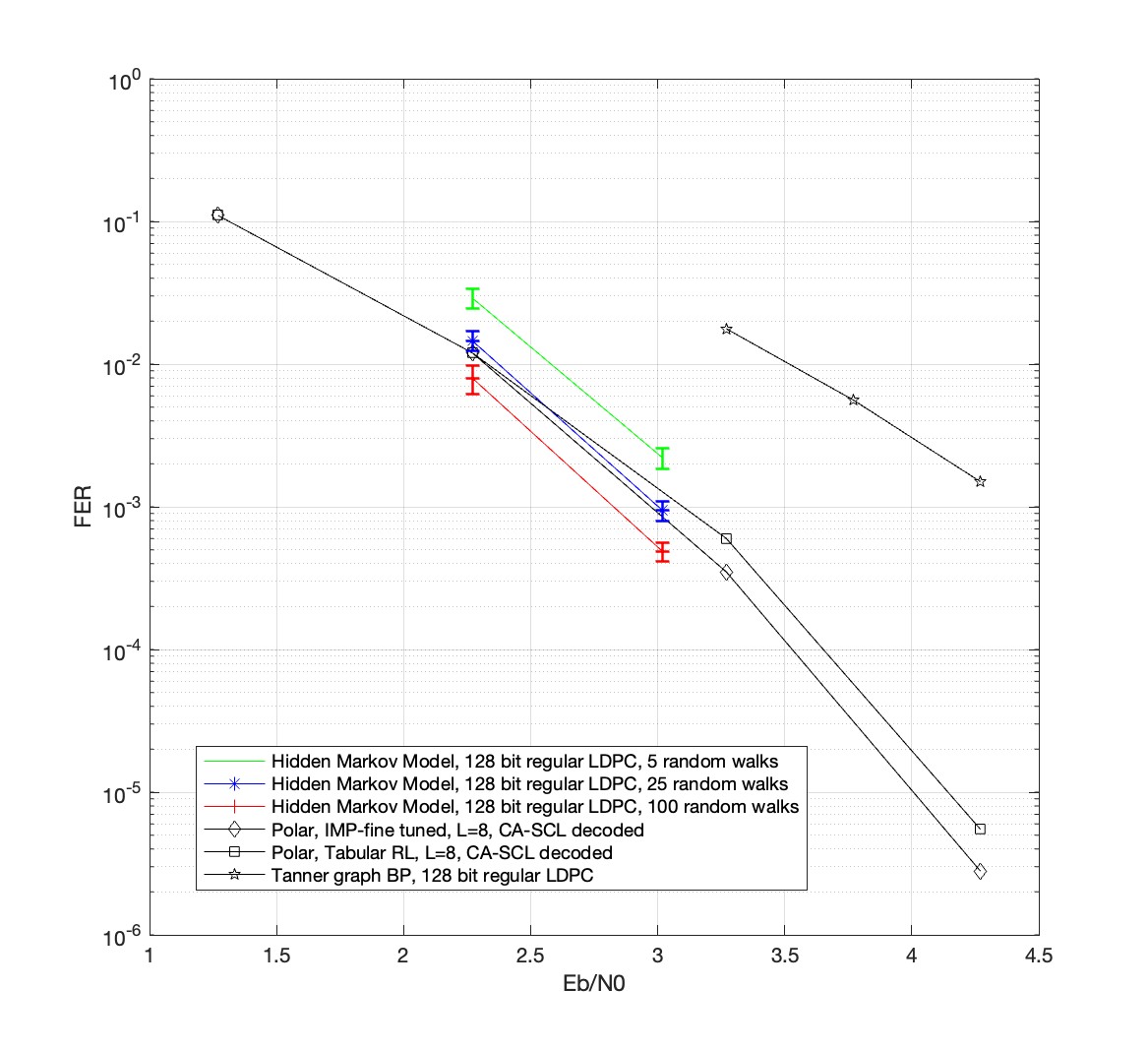}
\caption{The FER performance for a regular LDPC code under the proposed HMM decoder, compared to a Polar code optimized through the IMP method.  The Polar code utilizes a list decoder that is aided by a CRC.  Results were obtained from  \cite{cioffi}.     }
\label{cioffi}
\end{figure}

Figure \ref{cioffi} shows how a regular LDPC code performs relative to the IMP designed Polar code, as well as the Tabular RL design presented in \cite{tabular}. 
The threshold of the LDPC code is similar to that of the $L=8$ Polar code.   It could well be that larger $L$ values for the Polar code will improve the FER of the Polar code, but for the results presented in \cite{cioffi} the regular LDPC code under the proposed HMM decoder does not show a loss in sensitivity compared to the Polar code. This is a new result, confirming that for short frames, a low-complexity LDPC code under the proposed HMM decoder presents an alternative to Polar codes where a complex decoder is required. 

Figure \ref{cioffi} also shows that by increasing the number of random walks the threshold can be improved.  Within the uncertainties shown, the $100$ random walk results are significantly better than those of the $5$ and $25$ random walk cases.  However, it is clear that the HMM decoder performs well even with a small number of random walks.  

Note that the threshold and gradient of the FER plot based on the HMM decoder is significantly better than that of the Tanner graph under BP.  As commented in the previous section, this indicates that the HMM decoder is able to better reveal the true potential of the regular LDPC code. 

\section{Conclusions} \label{concluded}

The paper proposed a novel iterative decoder for an LDPC code. The decoder is based on a first-order Hidden Markov Model (HMM).  The paper demonstrated that the HMM employing compound hidden states presents a natural framework to enforce a finite number of parity checks contained by the parity check matrix $\mathbf H$. 

Each iteration contains a forward-backward message passing estimation of the hidden states, which represents a random walk through the coded bits.  As decision feedback is employed after each iteration to update the evidence, inference on the HMM chain is sub-optimal. 

Based on Monte Carlo simulations, the paper showed that the proposed iterative HMM decoder significantly improves the decoding thresholds that have been reported in the literature for short regular LDPC codes\footnote{Based on a Tanner graph under belief propagation.}. Moreover, the paper demonstrates that the regular LDPC code threshold is improved to the point where it rivals that of Polar codes of the same length.

Given the high complexity of the List-CRC decoder for the Polar code, and the low complexity of the message passing hidden state estimator on an HMM chain, a regular LDPC code under the iterative HMM decoder offers the system designer a new option when selecting a code. 

Irregular LDPC codes were not studied, but offer potential further improvement to the decoding threshold. 

\appendices

\section{The forward-backward smoothing estimator for the first order HMM}
Denote the emission model (observed data and hidden state) as $P(\mathbf E_k=e_k \big | \mathbf X_k)$, where $e_k$ denotes the observed value at the $k$-th evidence value.  The forward-backward smoothing algorithm can be compactly formulated based on matrix algebra. Each state assume $Q=4$ values denoted as $\small{\mathbf X_k = \{ x_1,x_2,\cdots,x_Q \}}$. 

An implementation of the MAP detector based on the forward-backward messaging algorithm, in terms of matrix operations, is as follows:
\begin{enumerate}
\item  Let the state transition model given by
\begin{equation}
P(\mathbf X_{k} \big| \mathbf X_{k-1} )
\end{equation}
is represented by a matrix $\mathbf T$.  
\begin{eqnarray} \small 
 \left[ \begin{array}{ccccccccccc}
{ P(x_k=x_1 \big| x_{k-1}=x_1) }   ~~\cdots  P(x_k=x_Q \big| x_{k-1} = x_1)   \\
P(x_k=x_1 \big| x_{k-1}=x_2)   ~~\cdots  P(x_k=x_Q \big| x_{k-1} = x_2)   \\
\hspace{0mm} \vdots  \hspace{20mm} \vdots  \hspace{0mm}  \\
P(x_k=x_1 \big| x_{k-1}=x_Q)   ~~\cdots  P(x_k=x_Q \big| x_{k-1} = x_Q)   \\
\end{array} \right] \nonumber
 \end{eqnarray}
 \item The observation (evidence) model given by 
 \begin{equation}
 P(e_k \big| \mathbf X_k)
 \end{equation}
 where $e_k$ is the value obtained from the evidence terminal (sensor),   is represented as a matrix $\mathbf O_k$ given by 
 \begin{eqnarray} \small 
 \left[ \begin{array}{ccccccccccccccccccccccccccccccc}
P(e_k \big| x_{k}=x_1)   & 0 & \cdots & 0   \\
0   & P(e_k \big| x_{k}=x_2) & \cdots & 0   \\
\hspace{0mm} \vdots & \hspace{0mm} \vdots & \ddots & \hspace{0mm} \vdots \\
0  & 0 & \cdots & P(e_k \big| x_{k} = x_Q)   \\
\end{array} \right] \nonumber
 \end{eqnarray}
 \item Represent the forward message column vector as $\mathbf z$.  The iterative forward message is given by 
 \begin{equation} \label{forward1}
 \mathbf z^{k+1} = \beta \mathbf O_{k+1}  \mathbf T^\dagger \mathbf z^{k}
 \end{equation}
and the backward message $\mathbf b$ is given recursively by 
\begin{equation} \label{back1}
\mathbf b^{k+1} =  \mathbf{T O}_{k+1}  \mathbf b^{k+2}.
\end{equation}
\item The probability distribution given the evidence, can be written as  
\begin{equation} \label{combine}
P(\mathbf X_k \big| \mathbf e_{1:N}) = \beta \mathbf z_k \otimes  \mathbf b_{k+1}.
\end{equation}
\end{enumerate}
The operator $\otimes$ implies point-by-point multiplication, and $\beta$ is set so that probabilities are normalized.

\bibliography{references}
\end{document}